# Enrichment of deeply penetrating waves in disordered media

Wonjun Choi[1], Moonseok Kim[1], Donggyu Kim[1], Changhyeong Yoon[1], Christopher Fang-Yen[1,2],

Q-Han Park[1] and Wonshik Choi[1,*]

[1]*Department of Physics, Korea University, Seoul 136-701, Korea*

[2]*Department of Bioengineering, University of Pennsylvania, Philadelphia,*

*Pennsylvania 19104, USA*

**Waves incident to a highly scattering medium are incapable of penetrating deep into the medium due to the diffusion process induced by multiple scattering[1,2]. This poses a fundamental limitation to optically imaging, sensing, and manipulating targets embedded in opaque scattering layers such as biological tissues[2,3]. One strategy for mitigating the shallow wave penetration is to exploit eigenmodes with anomalously high transmittance existing in any disordered medium. When waves are coupled to these eigenmodes, strong constructive wave interference enhances deeply penetrating waves [4-9]. However, finding such eigenmodes has been a challenging task due to the complexity of disordered media[5,8,10]. In this Letter, we present an iterative wavefront control method that selectively enriches the coupling of incident beam to high-transmission eigenmodes. Specifically, we refined the high-transmission eigenmodes from an arbitrary initial wave by either maximizing transmitted wave intensity or minimizing reflected wave intensity. Using the proposed method, we achieved more than a factor of 3 increases in light transmission through a scattering medium exhibiting hundreds of scattering events. Our approach is readily applicable to *in vivo* applications in which only the detection of reflected waves is available. Enhancing light penetration will lead to improving the working depth of optical imaging and treatment techniques[2,11,12].**

Disordered media such as human skin, white powders and fogs, although they are not strong absorbers, appear to be opaque. This is due to a large number of scattering events occurring inside these media, which spoil the directionality of wave propagation and attenuate wave transmission. Similar to the electric resistors in which scattering of free electrons with fixed atoms causes electrical resistance, the disordered media cause resistance to the energy flow[1,13]. Considering that the scattering is random process, it appears to be almost impossible to clear up this resistance. However, recent studies have shown that wave propagation in the disordered media can be controlled by exploiting wave interference[5,8,9,14-18]. In fact, shaping incident wave into a specific pattern can lead to its coupling to high-transmission eigenmodes, thereby inducing strong constructive interference at the opposite side of the disordered medium and increasing wave transmission. However, finding a proper incident beam pattern is a substantial task due to the complexity of a disordered medium. In general, one needs to obtain medium's input-output response in full details by recording thousands of transmission images, during which the medium can be affected by external perturbations. Moreover, the recording of transmission images is not feasible for almost all the practical applications because detector sensors cannot be located on the opposite side of the medium to the incident wave.

We consider an iterative feedback control as an alternative that can search for the proper incident pattern from an arbitrary starting pattern. If successfully implemented, it will remove the necessity of lengthy image recording. One possible control method implemented so far is to segment incident wave into two parts, parts I and II as shown in Fig. 1(c), and to shift the relative phase, $\Delta\phi$, of the part II with respect to the part I. As we increase $\Delta\phi$, the intensity of the transmitted wave at a single point in the detector sensor (let's say a point A indicated as a gray dot) exhibits sinusoidal modulation due to the wave interference. Therefore, one can choose a particular value of $\Delta\phi$ to maximize the intensity at the point[5]. On the other hand, the intensity at another point (point B, for instance) also exhibits sinusoidal modulation, but its maximum occurs at a different value of $\Delta\phi$. This implies that it is not straightforward to maximize the total intensity across the entire detector sensor.

According to our theoretical study, we found that the control of $\Delta\phi$ can increase total transmission due to the following two reasons. At first, the total intensity also exhibits sinusoidal modulation with the increase of $\Delta\phi$. Because the intensity modulation at each and every pixel is sinusoidal with the same period, summation of intensity across all detection points leads to the sinusoidal modulation. Therefore, we can still find $\Delta\phi = \Delta\phi^m$ that maximizes the total intensity. Then the question arises whether this maximization is a meaningful step, which requires the introduction of the second reason - Assigning $\Delta\phi^m$ leads to the enrichment of coupling to high-transmission eigenmodes. In order to understand this second reason, we need to view wave transmission in terms of eigenmodes. According to random matrix theory (RMT), in any disordered medium there exist a number of eigenmodes whose transmittance varies in a wide range[19-21]. One can sort eigenmodes in the descending order of their amplitude transmittance and assign an eigenmode index, $i$, to the amplitude transmittance of each eigenmode, $\tau_i$, such that $\tau_i \geq \tau_j$ for $i < j$. When an arbitrary wave, $E^{in}$, is sent to a disordered medium, it is decomposed into various eigenmodes.

$$E^{in} = \sum_{i=1}^{N} c_i v_i. \tag{1}$$

Here, $v_i$ is the eigenmode at the input plane, $c_i$ is the scalar product between $v_i^*$ and $E^{in}$, and $N$ is the total number of the eigenmodes. After the incident wave is transmitted through the medium, the coefficient of each eigenmode, $c_i'$, is multiplied by its amplitude transmittance, i.e. $c_i' = \tau_i c_i$. Therefore, the transmitted wave is preferentially occupied by the high-transmission eigenmodes in proportion to their transmittance (i. e. $|c_i'|^2 \propto \tau_i^2$).

The transmitted waves originating from both parts I and II are weighted sum of eigenmodes by their respective transmittance. Then, the modulation of total intensity due to the change in $\Delta\phi$ can now be interpreted as the summation of intensity modulations of individual eigenmodes, instead of the modulations at individual detector points. An important point in this eigenmode description is that the amplitude of sinusoidal modulation for high-transmission eigenmodes is larger than that of low-transmission ones. This is in strong contrast to previous description in which the intensity modulation at each pixel is independent. As a result, assigning the $\Delta\phi^m$ preferentially increases the intensity modulation of high-transmission eigenmodes. We have put this feedback control into an iterative loop to sequentially maximize the total transmission. We conducted numerical analyses to show that successive repetitions of this optimization process continue to enrich the high transmittance eigenmodes (See Supplementary Information for detailed theory). The important benefit of this approach lies in that it can *minimize* total reflection to enhance total transmission. This makes our approach readily applicable to the *in vivo* application.

The schematic diagram of our experiment is shown in Fig. 1(a). We used a He-Ne laser with the wavelength of 633 nm as a light source and illuminated the output beam from the laser to a spatial light modulator (SLM, LC2500R, Hamamatsu) in order to shape the pattern of the beam. We performed two separate modes of experiments, one in transmission and the other in reflection. In the transmission mode of experiment, we monitored the intensity of transmitted wave using a photodetector, PD1, and maximized its intensity by the feedback control of SLM. On the other hand, we monitored the reflected wave using a photodetector, PD2, in the reflection mode and minimized total reflected intensity by the control of SLM to enhance transmission.

We first performed transmission mode of experiment to explicitly prove the working principle of the feedback control. In order to choose and control parts I and II, we randomly divided the SLM pixels into two parts as shown in Fig. 1(b). Half of the pixels, colored blue, constitute part I and the other half, colored red, constitute part II. Initially, the phase value at each pixel was randomly chosen. The illumination beam was circular with a diameter of 11.4 $\mu m$. As a disordered medium, we used layers of $TiO_2$ particles with 10.9 % average transmittance and thickness of $8 \pm 2$ μm (See Methods for the sample preparation). An aperture and a polarizer were put in front of PD1, which reduced the measured transmittance to 1.55 %, in order to compare the feedback control method with the direct recording of a transmission matrix to be explained in Fig. 3. To make an overall phase shift $\Delta\phi$ for part II, the value $\Delta\phi$ was added to all the red pixels. The measured total intensity with the increase of $\Delta\phi$ indeed exhibited sinusoidal modulation (Fig. 2(a)). After assigning the first choice of $\Delta\phi^m$ to part II, we repeat the process using a new random choice of part I and II to further maximize the total transmission. When we applied the iterative feedback control algorithm, the total intensity of transmitted light indeed increased (Fig. 2(b)). Figures 2(c) and 2(d) show the images of transmitted wave before and after the feedback control, respectively, which clearly show that the total intensity was dramatically increased. The best experimental record of total transmission enhancement was a factor of 3.34 when compared with the transmittance of uncontrolled wave. Limiting factors in the amount of transmission increase were found to be the stability of the experimental setup.

In order to validate that the high-transmission eigenmodes were indeed enriched, we measured the transmission matrix of the same medium as done previously[8] and obtained all the eigenmodes present in the system. We then estimated the contribution of each eigenmode to the feedback-controlled incident wave. For this purpose, we calculated the cross-correlation between the incident wave and each eigenmode, and plotted the absolute square of the correlation as a function of the eigenmode index (Fig. 3(a)). The contribution of eigenmode with a smaller index (or larger transmittance) was increased by the feedback control, in good agreement with the theoretical prediction (Supplementary Information). We observed both a steady growth of high-transmission eigenmodes and a gradual decline of the small-transmittance ones (Fig. 3(b)) with increased number of iterations. These observations confirm the enrichment of high-transmission eigenmodes by our feedback control method. In fact, this enrichment is also supported by the observation that the contrast of total intensity modulation with $\Delta\phi$ was increased over the iterations (Fig. 2(a)). This indicates that the effective number of eigenmodes involved in the interference was reduced. As a short note, the previously reported single-point optimization method in which the detection area is minimally small is a limiting case of our method. So the relative transmission enhancement is typically a few times smaller than the proposed method (Supplementary Information).

The importance of our method lies in its applicability to the *in situ* control of light energy delivery. Most of the past studies addressing the issues of multiple light scattering in highly scattering media have dealt with the transmission mode of detection, in which a detector sensor is located on the opposite side of the medium from the incident wave[5,8,22,23]. In many practical applications, however, only the reflected waves can be recorded. Therefore, we decided to perform the reflection mode of experiment to control and reduce the intensity of the reflected waves. In the experimental setup shown in Fig. 1(a), we measured the total intensity of the reflected waves by PD2 while controlling the overall phase $\Delta\phi$ of segment II. We then found the $\Delta\phi$ that *minimizes* the total intensity of the reflected waves, and iteratively repeated the process. This feedback process preferentially induced the destructive interference of high-reflection eigenmodes such that low-reflection eigenmodes, i.e. high-transmission eigenmodes, were refined (See Supplementary Information for more detailed theory on minimization). As shown in Fig. 4(a), the reflected wave intensity decreases as the number of iterations is increased. Here, the same disordered medium used in the transmission mode of experiment was used. While minimizing the reflected wave intensity, we simultaneously measured the intensity of the transmitted wave by PD1 and observed that the transmission indeed increased. This validates that we can enrich high-transmission eigenmodes by the control of reflected waves. This is the first *in situ* demonstration of enhancing light energy delivery deep into a scattering medium. As a reference, we can compare the minimizing of the intensity at a single point of the reflected wave. While single-point optimization can enhance transmission to a certain degree, single-point minimization makes negligible contribution to reducing reflectance (See Supplementary Information). Therefore, in the reflection mode, the proposed method is the only possible feedback control approach developed to date.

To conclude, we have developed a simple and robust feedback control method that iteratively minimizes the reflected wave intensity and enhances light penetration through a disordered medium. Sensing, imaging, and manipulating a target hidden under highly scattering layers are universal problems where efficient illumination to the target is prerequisite condition. If the proposed method is combined with high-speed spatial light modulating devices such as digital micromirror devices, our method will readily find numerous important applications. For example, the proposed method can increase the working depth of optical imaging such as diffuse optical tomography[24] and photoacoustic tomography[25,26], and increase the treatment depth of phototherapies[11,12] and laser surgeries, and facilitate deep-tissue optical manipulations such as in optogenetics[27,28]. Since the methodology works for all types of waves, its applicability can be extended to microwave[29] and sound wave technologies[30].

**Methods Summary**

**Feedback control algorithm.** In either transmission or reflection mode of experiment, we scanned the overall phase $\Delta\phi$ of the part II of the SLM and measured total intensity at each phase. In this process, we chose four steps of $\Delta\phi$ at the interval of $\pi/2$ and recorded total intensities, $I(\Delta\phi = 0)$, $I(\pi/2)$, $I(\pi)$, and $I(3\pi/2)$. We then found coefficients A, B, and $\phi_{I,II}$ of a sinusoidal function $I(\Delta\phi) = A + B\cos(\Delta\phi + \phi_{I,II})$ that fits to the four measurements. We assigned $\Delta\phi = -\phi_{I,II}$ in the transmission mode of experiment, and $\Delta\phi = -\phi_{I,II} + \pi$ in the reflection mode. The current iterative process takes about 30 minutes mainly due to the slow refresh rate of the SLM (10 Hz).

**Sample preparation**

In this experiment, $TiO_2$ particle (Sigma-Aldrich 204757) layers were used as disordered media. $TiO_2$ particles have high refractive index (2.58) and are almost free from absorption for the laser with 633 nm wavelength. In order to prepare relatively uniform layers, we made a solution of $TiO_2$ in ethanol and spread the layer on the slide glass (or cover glass) by using air spray. The thickness of the $TiO_2$ layer measured by atomic force microscopy was $8 \pm 2$ μm and the transport mean free path found from the relation between the transmission and the thickness of the layer was $0.5 \pm 0.2$ μm.

**Acknowledgements**

This research was supported by the Basic Science Research Program through the National Research Foundation of Korea (NRF) funded by the Ministry of Education, Science and Technology (2011-0016568), the National R&D Program for Cancer Control, the Ministry of Health & Welfare, South Korea (1120290).

**Supplementary Information**

Supplementary text and Figures S1-S8.

**Figure legends**

**Figure 1. Schematic diagram of experimental apparatus and the description of the feedback control. a**, Schematic experimental setup. The output beam from a He-Ne laser was reflected by a mirror (M) and subsequently by a beamsplitter (BS1), and illuminated SLM. The reflected wave by the SLM transmitted through BS1 and BS2 to illuminate a disordered medium (S). PD1 and PD2 recorded total intensity of transmitted and reflected waves, respectively, by the sample. **b**, Representative phase pattern written on SLM. Pixels colored in blue constitute part I, and those in red constitute part II. Color bar, phase in radians. **c**, Schematic diagram of feedback control. The incident wave was segmented into two parts indicated by I and II, and the second part is shifted in phase by $\Delta\phi$. A detector sensor (D) records the transmitted wave through a scattering sample (S). A gray point indicated by "A" is used for the single-point optimization, and the entire sensor area is used for total transmission optimization. **d**, Experimentally measured intensity at pixels A and B in **c** as a function of $\Delta\phi$ for a disordered medium with 10.9 % average transmittance.

**Figure 2. The transmission enhancement by a feedback control. a**, Experimentally measured total transmission as a function of $\Delta\phi$ at the iteration steps of 1, 101, 351, and 1601. **b**, Experimentally measured transmittance over the increase of the number feedback control iterations. **c** and **d**, The intensity images of the transmitted wave before and after optimization, respectively. Scale bar, 5 $\mu$m. Color bar indicating intensity in arbitrary units applies to both **c** and **d**.

**Figure 3. Enrichment of high-transmission eigenmodes in the transmission mode of experiment. a**, The eigenmode distribution of the incident wave before (green) and after (blue) feedback control optimization. **b**, The contribution to the incident wave of the first 7 eigenmodes (blue) and last 7 eigenmodes (green) vs. the number of feedback control iterations.

**Figure 4. Transmission enhancement by the reflection mode of feedback control. a**, Total reflectance with the increase of the number of iterations (blue) and total transmittance (green) simultaneously measured during the feedback control. **b** and **c**, Intensity images of the reflected wave before and after the feedback control. **d** and **e**, Intensity images of the transmitted wave before and after the feedback control. Scale bar, 10 $\mu$m. Color bar indicates intensity in arbitrary units.

Figure 1

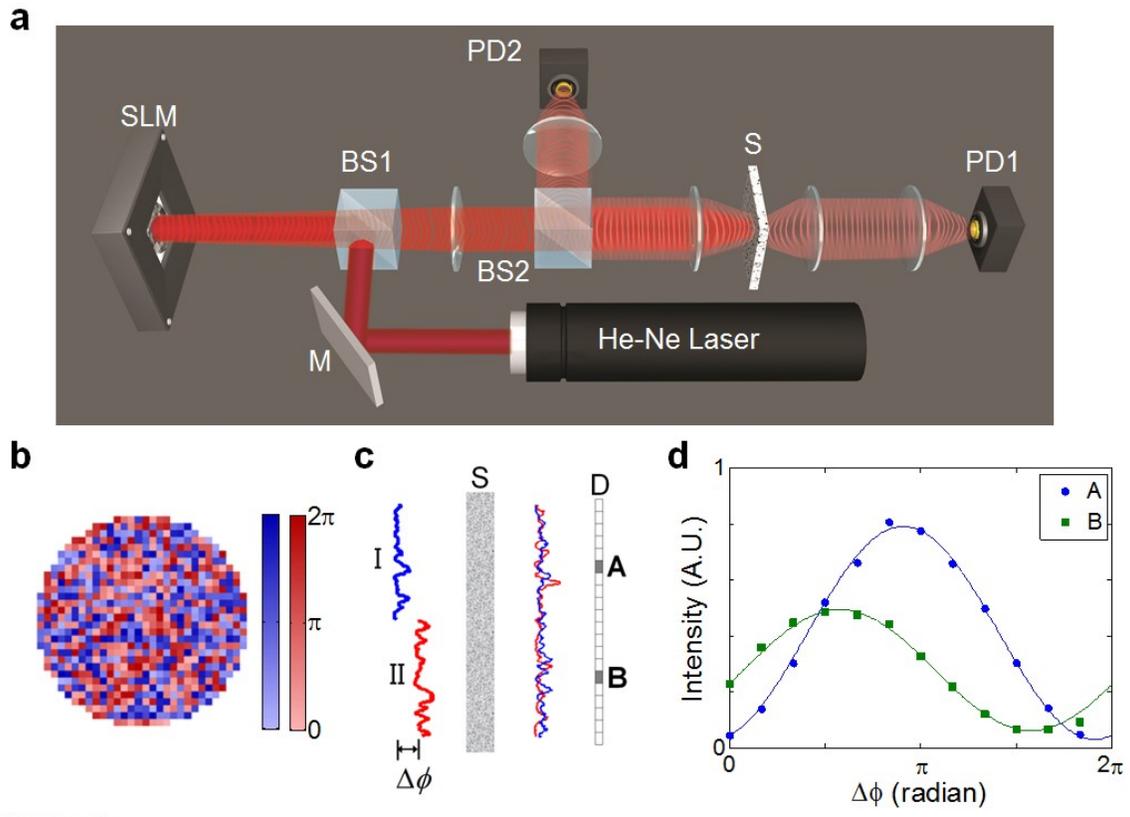

Figure 2

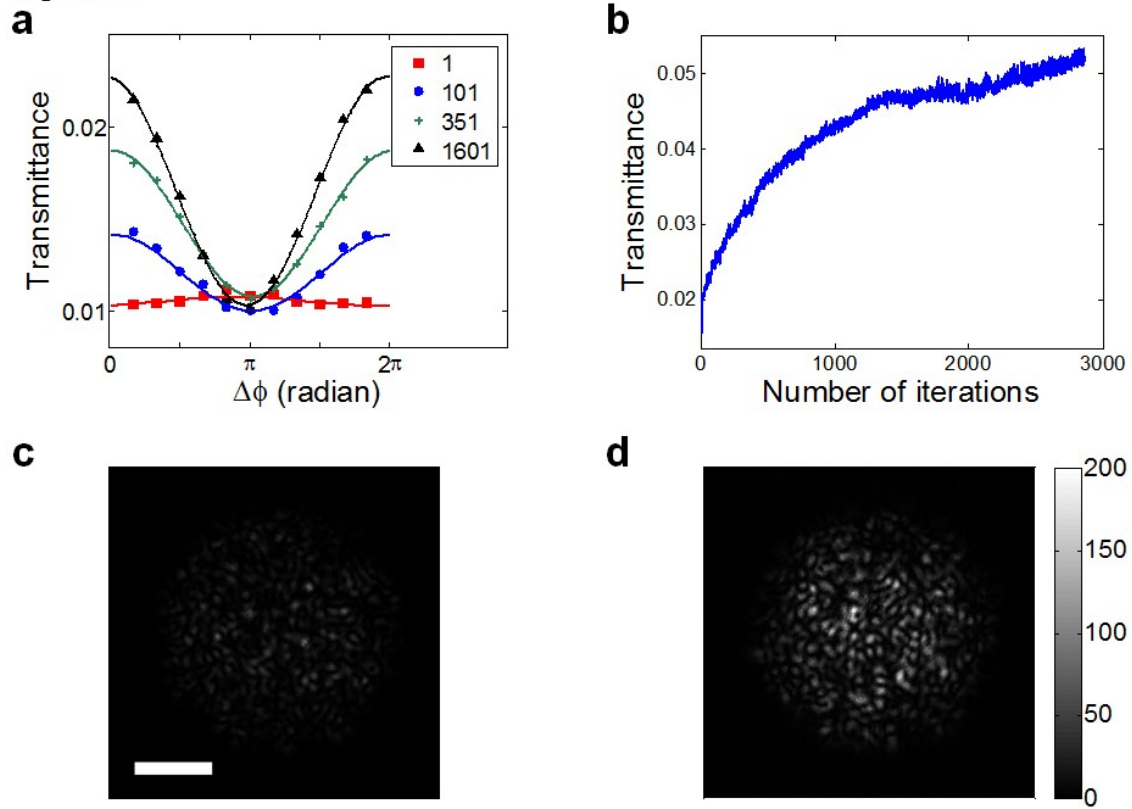

Figure 3

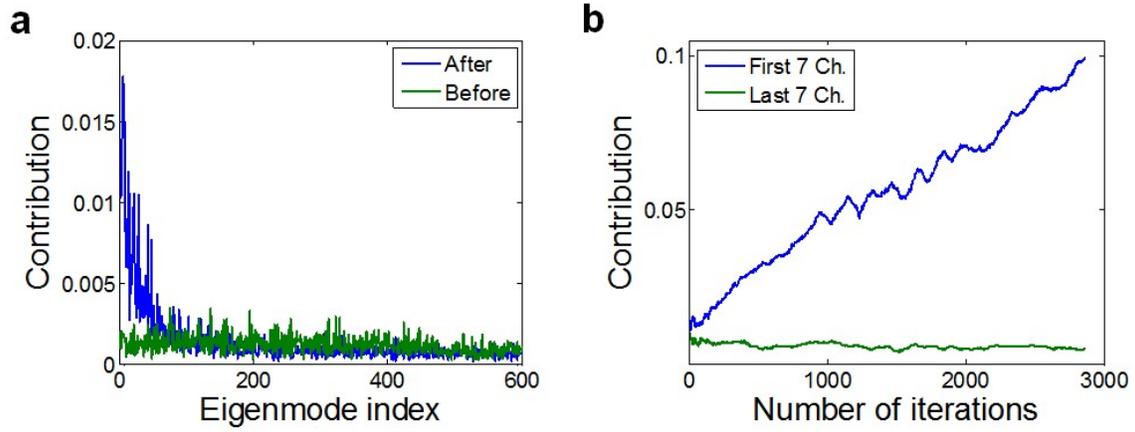

Figure 4

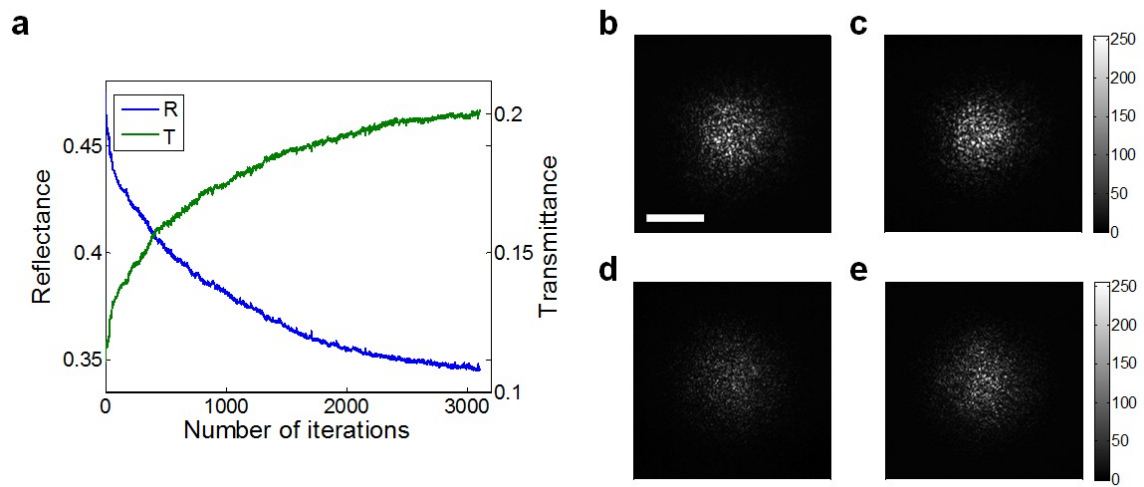

14